\documentstyle[epsfig]{aipproc}

\newcommand{\bfnabla}{\mbox{\boldmath $\nabla$}}
\newcommand{\dd}{{d}}
\newcommand{\tr}{{\rm tr}}
\newcommand{\Tr}{{\rm Tr}}

\newcommand{\thru}[1]{\mathrel{\mathop{#1\!\!\!/}}}

\newcommand{\bfp}{\mbox{\boldmath $p$}}

\newcommand{\bfx}{\mbox{\boldmath $x$}}

\newcommand{\bfD}{\mbox{\boldmath $D$}}
\newcommand{\bfDt}{{\tilde{\bfD}}}

\newcommand{\bfM}{\mbox{\boldmath $M$}}
\newcommand{\bfMt}{{\tilde{\bfM}}}

\newcommand{\bfY}{\mbox{\boldmath $Y$}}

\newcommand{\Ac}{{\cal A}}

\newcommand{\bra}{\langle}
\newcommand{\ket}{\rangle}

\begin{document}
\title{ Anomalies for Nonlocal Dirac Operators
}

\author{\underline{E. Ruiz Arriola}\footnote{Talk given at the
"Workshop on Hadron Physics", Coimbra (Portugal), 10-15
September 1999.} and L. L. Salcedo}
\address{
{~} \\
Departamento de F\'{\i}sica Moderna.
Universidad de Granada \\
E-18071 Granada, Spain}
\maketitle

\begin{abstract}
The anomalies of a very general class of non local
Dirac operators are computed using the $\zeta$-function definition of
the fermionic determinant and an asymmetric version of the 
Wigner transformation. For the axial anomaly all new terms
introduced by the non locality can be brought to the standard 
minimal Bardeen's form. Some extensions of the present techniques 
are also commented. 
\end{abstract}

\section*{Introduction}

Local field theories provide the commonly accepted setup  where the
implementation of space-time symmetries becomes rather simple. On the
other hand, effective theories are not necessarily local, although an
appropriate choice of degrees of freedom can make them almost local
\cite{We79}.
In low  energy QCD, light quarks and gluons are dressed by the 
interaction in a way that the effective theory looks highly nonlocal 
\cite{MP78}, and a kind of dynamical perturbation theory 
would be needed \cite{PS79}. In terms of pions and (heavy) nucleons 
the theory becomes weakly nonlocal and a chiral perturbation theory 
becomes of practical interest \cite{GL84}. In a Dyson-Schwinger
setting \cite{RW94} most information about such a non local theory comes
from the constraints imposed by the relevant Ward and Slavnov-Taylor
identities \cite{RW94}, perturbation theory to some finite order
and hadronic phenomenology. These approaches are necessary if 
one wants to know, for instance, about the momentum distribution of a 
quark in a hadron; neither perturbative QCD nor chiral perturbation theory   
can properly handle this problem. In this regard, anomalies are particulary
interesting because their existence is linked to a violation
of classical symmetries by high energy regulators although their
physical effect is formulated as a low energy theorem.  
At the one loop level, anomalies for nonlocal models have been previously 
discussed \cite{RC88,HT89,BR93} for some specific processes like e.g. 
$\pi^0 \to2 \gamma $, $ \gamma \to 3 \pi $ and $ 2 K \to 3 \pi $. We refer to
those works and \cite{Ri99} for further motivation. Rather than computing 
all specific processes one by one we prove that the new terms generated by the non locality can be subtracted by adding suitable counterterms. 

\section{The one loop effective action}

Our starting point is the effective action of Dirac fermions 
in the flat Euclidean space-time  endowed with internal degrees of freedom
 collectively referred to as ``flavor'', 
\begin{equation}
W(\bfD)=
-\log\int {\cal D}\bar\psi{\cal D}\psi \exp \left\{
-\int \dd^Dx\,\bar\psi(x)\bfD\psi(x)\right\}
=-\Tr\log\bfD\,.
\end{equation}
Here, $\Tr$ stands for trace over all degrees of freedom and $\bfD$ 
is the Dirac operator to be specified below. The definition of the fermion 
determinant requires some renormalization of the ultraviolet divergences. 
The (consistent) chiral anomaly is  defined as the variation of the 
effective action under infinitesimal chiral transformations, given by 
\begin{equation}
\psi(x) \to  e^{i\beta-i\alpha\gamma_5} \psi(x) \, , \qquad 
\bar \psi(x) \to  \bar \psi(x)  e^{-i\beta-i\alpha\gamma_5} 
\end{equation}
where $\alpha(x)$ and $\beta(x)$ are Hermitian matrices in flavor
space only, regarded as multiplicative operators on the fermionic
wave functions. The particular cases $\alpha=0$ and $\beta=0$
correspond to vector and axial transformations, respectively. This induces 
transformations of the Dirac operator 
\begin{equation}
\bfD \to e^{i\beta-i\alpha\gamma_5}\bfD e^{-i\beta-i\alpha\gamma_5}\,,
\end{equation}
which infinitesimally become  
\begin{equation}
\delta\bfD = \delta_V\bfD+ \delta_A\bfD=
[i\beta,\bfD]-\{i\alpha\gamma_5,\bfD\}\,.
\end{equation} 
Since we will be considering a $\zeta$-function renormalization of $W$ (see 
below), there will be no vector anomaly,
\begin{equation}
\delta_VW=0, \quad \delta_AW=\Ac_A\,.
\end{equation}
Correspondingly, the same current conservation formulas valid for the
local case can be written here,
\begin{eqnarray} 0 &=& \int\dd^Dx\,\langle\bar\psi(x)[i\beta,\bfD]\psi(x)\rangle_Q
\\
-\Ac_A &=&
\int\dd^Dx\,\langle\bar\psi(x)\{i\alpha\gamma_5,\bfD\}\psi(x)\rangle_Q \,.
\end{eqnarray}
(The symbol $\langle\ \rangle_Q$ stands for quantum vacuum expectation
value.) In particular the term $\gamma_\mu P_\mu$ in $\bfD$ in the
right-hand side yields, after integration by parts, the divergence of
the fermionic vector and axial currents whereas the other terms in
$\bfD$, local and non local, represent the explicit chiral symmetry
breaking due to the external fields. On the other hand, the left-hand
side shows the anomalous breaking of the axial current
conservation. 

The class of Dirac operators to be considered here is
\begin{equation}
\bfD=\bfD_L + \bfM \,.
\end{equation}
The term $\bfD_L$, the local component of $\bfD$, is a standard Dirac
operator
\begin{equation}
\bfD_L=  \gamma_\mu P_\mu+\bfY
\end{equation}
We will follow the conventions of \cite{Sa96,Ru99}),
$P_\mu=i\partial_\mu$ and $\bfY$ is an
arbitrary matrix-valued function in flavor and Dirac spaces. $\bfY$ is
a function of the position operators
$X_\mu$, defined by $X_\mu\psi(x)=x_\mu\psi(x)$, so that $\bfY$ is a
multiplicative operator in the Hilbert space of fermions. The term
$\bfM$ is a purely non local
\footnote{
$\bfM$ is softer in the ultraviolet
sector than any multiplicative operator, that is, the distribution
$\bfM(x,y)$ is less singular than the Dirac delta $\delta(x-y)$.}
, more precisely bilocal, operator
also with arbitrary structure in flavor and Dirac spaces,
\begin{equation}
(\bfY\psi)(x)=\bfY(x)\psi(x) \, , \qquad
(\bfM\psi)(x)=\int \dd^Dy\bfM(x,y)\psi(y)\,.
\label{eq:4}
\end{equation}
More restrictive assumptions on $\bfM$ are spelled out in ref.~\cite{Ru99}.
For the purpose of doing detailed calculations we will assume that the
non local operator $\bfM$ admits an expansion in inverse powers of
$P_\mu$ for large $P_\mu$ of the form
\begin{equation}
\bfM=\bfM_\mu\frac{P_\mu}{P^2}+
\bfM_{\mu\nu}\frac{P_\mu P_\nu}{P^4}+
\bfM_{\mu\nu\rho}\frac{P_\mu P_\nu P_\rho}{P^6}+
\cdots
\label{eq:12}
\end{equation}
The coefficients $\bfM_{\mu_1\dots\mu_n}$ are multiplicative operators
and they are completely symmetric under permutation of indices. For
convenience the $P_\mu$ has been put at the right
\footnote{This choice does not exhaust all possible non local
operators, but it is realistic
since it accommodates the operator product
expansion estimate of the quark self-energy $  \Sigma (p^2 ) \sim_{p^2
\to \infty} ( {\rm  log} p^2 )^{d-1} / p^2 $ with $d$ the anomalous
dimension of the quark
 condensate $\bar \psi \psi $ (see ref.\cite{Po76})}. The
better way to obtain the transformation
properties of $\bfM$ is by introducing a family of operators
associated to $\bfM$ as
\begin{equation}
\bfMt(p)= e^{ipX}\bfM e^{-ipX}
\end{equation}
where the momentum $p_\mu$ is just a constant c-number. Effectively,
$\bfMt(p)$ corresponds to make the replacement $P_\mu\to P_\mu+p_\mu$
in $\bfM$. The function $\bfMt(p)$ admits an expansion in inverse
powers of $p_\mu$ similar to that in eq.~(\ref{eq:12}), namely
\begin{equation}
\bfMt(p)=\bfMt_\mu\frac{p_\mu}{p^2}+ \bfMt_{\mu\nu}\frac{p_\mu
p_\nu}{p^4}+ \cdots\,.
\label{eq:15}
\end{equation}


We will adopt the
$\zeta$-function renormalization prescription combined with an
asymmetric Wigner transformation. This method, as well as
several of its applications, is presented in great detail in
\cite{Sa96}. The $\zeta$-function effective action is given
by~\cite{Se67,Ha77}
\begin{equation}
W(\bfD) = -\Tr\log \bfD =
-\frac{\dd}{\dd s} \Tr\left(\bfD^s\right)_{s=0}\,,
\end{equation}
where $s=0$ is to be understood as an analytical extension on $s$ from
the ultraviolet convergent region ${\rm Re}(s)<-D$.
The operator $\bfD^s$ can be obtained from
\begin{equation}
\bfD^s = -\int_\Gamma\frac{\dd z}{2\pi i} \frac{z^s}{\bfD-z}
\end{equation}
where the integration path $\Gamma$ starts at $-\infty$, follows the
real negative axis, encircles the origin $z=0$ clockwise and goes back
to $-\infty$ (see figure 1). The key point is
that for sufficiently negative $s$ there are no ultraviolet
divergences and formal operations become justified. By construction, the
$\zeta$-function renormalized effective action is invariant under all
symmetry transformations associated to similarity transformations of
$\bfD$, thus in particular it is vector gauge invariant. On the other
hand, the variable $z$ plays the role of a mass and hence breaks
explicitly both chiral and scale invariance.

\begin{figure}
\centerline{
\epsfig{figure=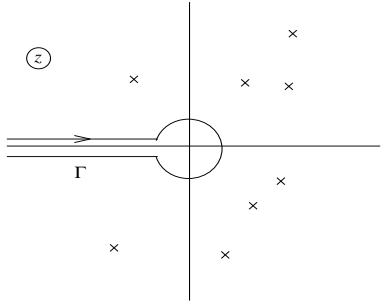,height=4cm,width=5cm}}
\caption{ \sf The $\Gamma$ contour in the complex $z$-plane. Crosses
represent isolated eigenvalues of $\bfD$.} 
\end{figure}

The operator $(\bfD-z)^{-1}$  can be conveniently
expressed by means of an asymmetric version of the Wigner
representation \cite{Wi32}.  For any operator $A$, let
\begin{equation}
A(x,p) =
\int\dd^Dy\,e^{iyp}\bra x|A|x-y\ket = {\bra x|A|p\ket\over
\bra x|p\ket} 
\end{equation}
be its (asymmetric) Wigner representation.  $|p\ket$ is
the momentum eigenstate with $\bra x|p\ket = e^{-ixp}$.
From this definition
\begin{equation}
\bra x|A|x\ket = \int{\dd^Dp\over(2\pi)^D}
A(x,p)\,,\quad \Tr A = \int{\dd^Dx\dd^Dp\over(2\pi)^D} \tr A(x,p)\,,
\end{equation}
where $\tr$ acts on internal and Dirac spinor degrees of freedom only, and the
product of two operators satisfy the following formula
\begin{equation}
(AB)(x,p) =
\exp(i\partial_p^A\cdot\partial_x^B)A(x,p)B(x,p)\,, 
\label{eq:pro}
\end{equation}
where $\partial^A_p$ acts only on the $p$-dependence in $A(x,p)$ and
$\partial_x^B$ on the $x$-dependence in $B(x,p)$.
Let the propagator or resolvent of $\bfD$, be
$G(z)=(\bfD -z)^{-1} $
and $G(x,p;z)$ its Wigner representation. Applying Eq.\ref{eq:pro}  
to $(\bfD-z)G=1$ one obtains 
\begin{equation}
G(x,p;z) = \langle x|(\thru p+\bfD-z)^{-1} |0\rangle 
\end{equation}
where $|0\rangle$ is the state of zero momentum, $\langle x|0\rangle = 1$.
In practice this implies that $i\partial_\mu$ derivates every
$x$ dependence at is right, until it annihilates $|0\rangle$, 
$i \partial_\mu |0\rangle=0$. This method is very efficient for a derivative
expansion for it computes directly, that is non recursively, each of the terms
\cite{Sa96}.

The definition given for $A(x,p)$ is not gauge covariant 
because $|p\ket$ is not. We will consider only local objects of the form $\bra
x| f(\bfD)|x\ket $ as given by the formula
$$ \bra x| f(\bfD)|x\ket = -\int {\dd^Dp\over (2\pi)^D} \int_\Gamma
{\dd z\over 2\pi i} f(z) G(x,p;z) \eqno(3.7) $$
We will assume that the function $f(z)$ is sufficiently convergent at
infinity or else that it can be obtained as a suitable analytical extrapolation
from a parametric family $f(z,s)$ in the variable $s$.
In either case the integration over $z$ should be performed in the
first place, to yield the Wigner representation of the operator $f(\bfD)$.
Afterwards, the $p$ integration is carried out, corresponding to
take the diagonal matrix elements of $\bra y|f(\bfD)|x\ket$, hence restoring
gauge covariance. This obviously means that the gauge breaking piece 
of $G(x,p;z)$ is a total derivative in the momentum variable. Recently, 
a method has been developped \cite{BP98} where this total derivative is, by 
construction, gauge invariant.

\section{ANOMALIES}


Because the chiral transformations are local, both $\bfD_L$ and $\bfM$
transform covariantly separately, that is,
\begin{equation}
\delta\bfD_L = [i\beta,\bfD_L]-\{i\alpha\gamma_5,\bfD_L\}\,, \quad
\delta\bfM = [i\beta,\bfM]-\{i\alpha\gamma_5,\bfM\}\,.
\end{equation}
Note that the bilocal structure of $M$ implies that local factors at
each side of the operator are taken at different points, i.e.
$ \bfM(x,x') \to \
         e^{i\beta(x)-i\alpha(x)\gamma_5} \bfM(x,x')
e^{-i\beta(x')-i\alpha(x') \gamma_5}$.

The two lowest coefficients are given by
\begin{equation}
\bfMt_\mu = \bfM_\mu \,, \qquad \bfMt_{\mu\nu} =
\bfM_{\mu\nu}+t_{\mu\nu\rho\sigma}\bfM_\rho P_\sigma \,,
\end{equation}
where we have introduced $t_{\mu\nu\rho\sigma}=
\delta_{\mu\nu}\delta_{\rho\sigma} -
\delta_{\mu\rho}\delta_{\nu\sigma}-\delta_{\mu\sigma}\delta_{\nu\rho}$.
It should be noted that the coefficients $\bfMt_{\mu_1\dots\mu_n}$ are
not multiplicative operators. One useful property of $\bfMt(p)$ is
that it transforms covariantly under chiral transformations. Indeed,
if $\bfM_\Omega=\Omega_1\bfM\Omega_2$ for two multiplicative operators
$\Omega_{1,2}$,
\begin{equation}
\bfMt_\Omega(p)= e^{ipX}\bfM_\Omega e^{-ipX}
=\Omega_1\bfMt(p)\Omega_2\,.
\end{equation}
As a consequence, the coefficients are also chiral covariant
\begin{equation}
\delta\bfMt_{\mu_1\dots\mu_n} = [i\beta,\bfMt_{\mu_1\dots\mu_n}]
-\{i\alpha\gamma_5,\bfMt_{\mu_1\dots\mu_n}\}\,.
\end{equation}
From here it is immediate to derive the transformation of the
original coefficients $\bfM_{\mu_1\dots\mu_n}$. For the
two lowest order coefficients one finds
\begin{eqnarray}
\delta\bfM_\mu &=&
[i\beta,\bfM_\mu]-\{i\alpha\gamma_5,\bfM_\mu\}\,, \\
\delta\bfM_{\mu\nu} &=&
[i\beta,\bfM_{\mu\nu}]-\{i\alpha\gamma_5,\bfM_{\mu\nu}\}
+t_{\mu\nu\rho\sigma}
\bfM_\rho(\partial_\sigma\beta+\partial_\sigma\alpha\gamma_5) \,. \nonumber
\end{eqnarray}
In general, the variation of each coefficient involves those of lower
order. 

The scale transformation $\psi(x)\to
e^{-\alpha_S(D-1)/2}\psi(e^{-\alpha_S}x)$ induces 
the corresponding transformation in $\bfD$, namely,
\begin{equation}
\bfY(x)\to e^{-\alpha_S}\bfY(e^{-\alpha_S}x)\,,\quad
\bfM_{\mu_1\dots\mu_n}(x) \to
e^{-\alpha_S(n+1)}\bfM_{\mu_1\dots\mu_n}(e^{-\alpha_S}x)\,.
\end{equation}
Infinitesimally, it implies 
\begin{equation}
\delta_S\bfD= -\alpha_S(\bfD-i[X_\mu P_\mu,\bfD])\,.
\end{equation}

\subsection{ Chiral anomaly } 

Due to the $\zeta$ regularization the chiral anomaly 
becomes an axial anomaly, 
\begin{equation}
\Ac_A = \Tr\left(2i\alpha\gamma_5\bfD^s\right)_{s=0}\,.
\end{equation}
Using the Wigner transformation technique~\cite{Sa96},
the anomaly can be written as (a similar expression holds for the
effective action)
\begin{equation}
\Ac_A = -\int\frac{\dd^Dp}{(2\pi)^D}\int_\Gamma\frac{\dd z}{2\pi i}z^s
\tr\langle 0|2i\alpha\gamma_5\frac{1}{\bfDt(p)-z}|0\rangle\Big|_{s=0}\,.
\end{equation}
Here $\tr$ stands for trace over Dirac and flavor degrees of freedom,
$|0\rangle$ is the zero momentum state normalized as $\langle
x|0\rangle=1$, thus $P_\mu|0\rangle=\langle0|P_\mu=0$. Further
\begin{equation}
\bfDt(p)= e^{ipX}\bfD e^{-ipX} = \thru{p}+\bfD_L+\bfMt(p)\,.
\end{equation}
The integration over $z$ should be performed first, since it defines
the operator $\bfD^s$, then the integral over $p$ which corresponds to
take the trace over space-time degrees of freedom and finally $s$ is
to be analytically extended to $s=0$. 
The simplest way to proceed is
to introduce a mass term, i.e., to apply the formula to the Dirac
operator $\bfD+m$ and then make an expansion in powers of
$\bfD_L+\bfMt(p)$, letting $m\to 0$ at the end. In this way the
following expression is derived
\begin{equation}
\Ac_A = \sum_{N \ge 0}
\int\frac{\dd^Dp}{(2\pi)^D}
\int_\Gamma\frac{\dd z}{2\pi i}z^s
\tr\langle 0|2i\alpha\gamma_5
\frac{(\bfD_L+\bfMt(p))\left((\thru{p}+z-m)(\bfD_L+\bfMt(p))\right)^N}
{(p^2+(z-m)^2)^{N+1}}
|0\rangle\Big|_{s=0, m=0}\,.
\end{equation}
Simplification has been achieved by using the cyclic property for the
trace in Dirac space.

After an angular average over $p_\mu$, the indicated integrals on
$p_\mu$ and $z$ can be carried out directly with the integral $I_1$
given in \cite{Sa96}. The result for the four dimensional chiral anomaly
is
\begin{eqnarray}
{\cal A}_A &=& \Big\langle 2i\alpha\gamma_5 \Big[
\frac{1}{2}\bfD_L^4 + \frac{1}{12}\bfD_L\{\gamma_\mu,\bfD_L\}^2\bfD_L
+\frac{1}{4}\bfMt_\mu^2 +\frac{1}{4}\{\bfD_L,\bfMt_{\mu\mu}\}
\nonumber\\ &&
+\frac{1}{8}\left(\bfMt_\mu\{\gamma_\mu,\bfD_L\}\bfD_L
+\bfD_L\{\gamma_\mu,\bfMt_\mu\}\bfD_L
+\bfD_L\{\gamma_\mu,\bfD_L\}\bfMt_\mu \right)
\Big] \Big\rangle
\label{eq:27}
\end{eqnarray}
The notation $\langle f\rangle$ stands for $
\langle f\rangle = \frac{1}{(4\pi)^{D/2}}\tr\langle 0|f(X)|0\rangle \,.$
Note that, even for non local Dirac operators, the anomaly is a
local polynomial of dimension $D$ constructed with $P_\mu$ and the
external fields $\bfY$ and $\bfM_{\mu_1\dots\mu_n}$. This is a
general property of all anomalies since only ultraviolet divergent
terms can contribute to them. The expressions found for the anomaly
can be put in a more usual form, in terms of vector and axial fields,
scalar fields, etc~\footnote{See ref~\cite{Sa96} for an explicit
expression in the local case and the remarks of ref~\cite{Ru99} for the
non local case.}, but it is preferable to use compact notation.
Since the regularization preserves vector gauge invariance, the axial
anomaly is also invariant. In our expression for the anomaly, this is
a direct consequence of the operators there being
multiplicative. Indeed, any operator $f$ constructed with the gauge
covariant blocks $\bfD_L$ and $\bfMt_{\mu_1\dots\mu_n}$ is also
covariant, i.e., $f\to \Omega f\Omega^{-1}$. If in addition $f$ is
multiplicative $\langle f\rangle$ is invariant. Note that
$\langle\,\rangle$ is not a trace and so the cyclic property does not
hold for arbitrary non multiplicative operators.

\subsection{Trace anomaly}\label{sec:4}

The corresponding trace anomaly, within the $\zeta$-function method
is~\cite{Sa96}
\begin{equation}
{\cal A}_S= \delta_S W= \alpha_S\Tr(\bfD^s)_{s=0}\,.
\end{equation}
The calculation is entirely similar to that of the axial anomaly, yielding 
\begin{eqnarray}
{\cal A}_S &=& \alpha_S\Big\langle
{1\over 2}\bfD_L^4 +
{1\over 12}(\bfD_L^2\{\gamma_\mu,\bfD_L\}^2+
\{\gamma_\mu,\bfD_L\}\bfD_L^2\{\gamma_\mu,\bfD_L\}
+\{\gamma_\mu,\bfD_L\}^2\bfD_L^2)
  \nonumber\\ &&
+ {1\over 96} (\{\gamma_\mu,\bfD_L\}^2\{\gamma_\nu,\bfD_L\}^2
+ (\{\gamma_\mu,\bfD_L\}\{\gamma_\nu,\bfD_L\})^2
+ \{\gamma_\mu,\bfD_L\}\{\gamma_\nu,\bfD_L\}^2\{\gamma_\mu,\bfD_L\})
  \nonumber\\ &&
+\frac{1}{12}(
 \gamma_\mu\bfMt_\mu\bfD_L^2
+\gamma_\mu\bfD_L^2\bfMt_\mu
+\gamma_\mu\bfD_L\bfMt_\mu\bfD_L
)
  \nonumber\\ &&
-\frac{1}{24}(
 \bfMt_\mu\gamma_\mu\bfD_L^2
+\bfD_L^2\gamma_\mu\bfMt_\mu
+\bfD_L\{\gamma_\mu,\bfMt_\mu\}\bfD_L
+\bfD_L\gamma_\mu\bfD_L\bfMt_\mu
+\bfMt_\mu\bfD_L\gamma_\mu\bfD_L
)
  \nonumber\\ &&
+\delta_{\mu\nu\alpha\beta}\Big(
\frac{1}{36}(
 \gamma_\mu\bfMt_\nu\gamma_\alpha\bfD_L\gamma_\beta\bfD_L
+\gamma_\mu\bfD_L\gamma_\alpha\bfMt_\nu\gamma_\beta\bfD_L
+\gamma_\mu\bfD_L\gamma_\alpha\bfD_L\gamma_\beta\bfMt_\nu
)
  \nonumber\\ &&
+\frac{1}{24}\gamma_\mu\bfMt_\nu\gamma_\alpha\bfMt_\beta
+\frac{1}{24}\{\gamma_\alpha\bfMt_{\mu\nu},\gamma_\beta\bfD_L\}
+\frac{1}{12}\gamma_\mu\bfMt_{\nu\alpha\beta}
\Big)
\Big\rangle\,.
\end{eqnarray}
Where $\delta_{\mu\nu\alpha\beta}=
\delta_{\mu\nu}\delta_{\alpha\beta}+\delta_{\mu\alpha}\delta_{\nu\beta}
+\delta_{\mu\beta}\delta_{\alpha\nu}$. The result is again a local
polynomial of dimension $D$ in the external fields and their
derivatives. Unlike the axial case, the coefficients $\bfM_{\mu\nu\alpha}$ 
in four dimensions do contribute to the scale anomaly.

Because scale and chiral transformations commute (in a properly defined
sense), the crossed variations $\delta_S{\cal A}_{V,A}$ and
$\delta_{V,A}{\cal A}_S$ coincide and they vanish since the axial
anomaly is scale invariant. Thus the scale anomaly must be chiral
invariant. The vector gauge invariance of the previous expressions is
easy to check noting that the operators inside $\langle\,\rangle$ are
multiplicative. Axial invariance is much more involved in general. 
In four dimensions it is relatively easy to check that the trace anomaly 
is axially invariant in the particular case of $\bfM_\mu=0$, which defines 
a class of operators invariant under chiral and
scale transformations.

\subsection{Counterterms and minimal form of the anomaly}\label{sec:3}

Presumably due to its topological connection~\cite{Al85}, the axial
anomaly is a very robust quantity. It is not affected by higher order
radiative corrections~\cite{AB69}, and remains unchanged at finite
temperature and density~\cite{Go94}. It gets no contributions from
scalar and pseudo scalar fields~\cite{Ba69}, tensor
fields~\cite{Cl83,Mi87,Sa96} or internal gauge fields, i.e,
transforming homogeneously under gauge transformations
~\cite{Bi93,Ru95}. In all known cases, the anomaly only affects the
imaginary part of the effective action in Euclidean space and only involves
vector and axial fields. The counter terms can always be chosen so that the
axial anomaly adopts the minimal or Bardeen's form~\cite{Ba69}. Not
surprisingly, the new terms introduced in the anomaly by the non local
component of the Dirac operator are also unessential, that is, they
can be removed by adding a suitable local and polynomial counter term
to the effective action. In other words, all new terms can be derived
as the axial variation of an action which is a polynomial constructed
with the external fields $\bfY$ and $\bfM_{\mu_1\dots\mu_n}$ and their
derivatives. The dimension of the polynomial can be at most
$D$.

The general proof that the anomaly can always be brought to
Bardeen's form has already been presented~\cite{Ru99}.
The actual construction of the counter terms can be done using the
method in ref.~\cite{Sa96} (see some further details in ref.~\cite{Ru98}). 
One interesting insight in the local case \cite{Sa96} is that the needed 
counterterms require not only the Dirac operator $\bfD$ but also its adjoint 
$\bfD^\dagger $ not related to the original theory.

The scale anomaly is already minimal. It can be modified by adding
polynomial counter terms of dimension smaller than $D$ but this would
add terms of the same type to the scale anomaly.

\section{Extension to finite temperature}

The Wigner transformation method combined with the $\zeta$-function
regularization has been further extended to the finite temperature
case in \cite{Salcedo:1998tg,Salcedo:1998sv}. As is well known, in the
imaginary time formulation of finite temperature field theory, the
field configurations are periodic or antiperiodic functions of the
Euclidean time for bosons and fermions respectively and thus the
frequency running in the fermion loop takes discrete values only,
$\omega_n=\pi(2n+1)T$ (where $T$ stands for the temperature and $n$ is
any integer) which are known as Matsubara frequencies. At finite
temperature, the trace of an operator
$f(x_0,\bfx;i\partial_0,i\bfnabla)$, acting on the Hilbert space of
$d+1$-dimensional fermions with possible internal degrees of freedom,
becomes
\begin{equation}
\Tr(f)= T\sum_n\int\frac{d^dp}{(2\pi)^d} 
\tr \langle 0|f(x_0,\bfx;\omega_n+i\partial_0,\bfp+i\bfnabla)|0\rangle\,.
\end{equation}
This formula generalizes that for zero temperature. Note that $f$ is a
periodic function of $x_0$, and $|0\rangle$ is the state with zero
momentum and energy normalized to $\langle x_0,\bfx|0\rangle=1$, and
thus it is periodic too.

In Ref.~\cite{Salcedo:1998tg} this method has been applied to compute
the anomalous component of the effective action of two- and
four-dimensional fermions at finite temperature in the presence of
arbitrary vector and axial gauge fields and scalar and pseudoscalar
fields on the chiral circle. The computation is carried out to leading
order in a suitable commutator which preserves chiral symmetry. As is
well known, at zero temperature the gauge Wess-Zumino-Witten (WZW)
action, which saturates the chiral anomaly, is the only leading
contribution in the anomalous sector; further terms must be Lorentz
and chiral invariant and they vanish identically unless they have more
gradients thereby being sub-leading terms. At finite temperature the
situation is different since Lorentz invariance is partially broken
and this allows to have new chiral invariant contributions\footnote{It
is well established that the chiral anomaly is
temperature independent (see e.g.~\cite{Go94}).} of the
same order as the WZW action. In particular these terms modify the
$\pi\to\gamma\gamma$ amplitude which is no longer determined by the
chiral anomaly~\cite{Pisarski:1997bq}. The calculation in
\cite{Salcedo:1998tg} confirms previous results
\cite{Pisarski:1996ne,Baier:1996vf} that this amplitude vanishes in a
chiral symmetric phase (see also \cite{Gupta:1997gn,Gelis:1998hb}).

In Ref.~\cite{Salcedo:1998sv} the same technique is applied to the
study of $2+1$-dimensional fermions at finite temperature in the
presence of arbitrary background gauge fields.  The use of the
$\zeta$-function regularization guarantees the gauge invariance of the
result under topologically small and large transformations. This has
allowed to solve a long standing puzzle, namely, the apparent
renormalization of the Chern-Simons coefficient at finite temperature,
which has been shown to be a perturbation theory
artifact~\cite{Dunne:1997yb}. In Ref.~\cite{Salcedo:1998sv} all
ultraviolet divergent terms of the effective action, within a strict
gradient expansion, have been computed. The result preserves gauge and
parity symmetries (up to the standard temperature independent parity
anomaly) and display the correct $2\pi i$ multivaluation introduced by
the Chern-Simons term. The known exact result for massless fermions
~\cite{Alvarez-Gaume:1985nf} is also reproduced.

\section*{Acknowledgments}
One of us (E.R.A.) acknowledges the organizers for the stimulating atmosphere 
during the workshop.  
This work is supported in part by funds provided by the Spanish DGICYT
grant no. PB95-1204 and Junta de Andaluc\'{\i}a grant no. FQM0225.

\end{document}